\documentclass[prd,preprint,superscriptaddress,showpacs,tightenlines]{revtex4}

\usepackage{graphics}
\usepackage{epsfig}

\begin{document}

\preprint{MKPH-T-06-18}

\title{Axial, induced pseudoscalar, and pion-nucleon form factors
in manifestly Lorentz-invariant chiral perturbation theory}
\author{M.~R.~Schindler}
\affiliation{Institut f\"ur Kernphysik, Johannes
Gutenberg-Universit\"at, D-55099 Mainz, Germany}
\author{T.~Fuchs}
\affiliation{Institut f\"ur Kernphysik, Johannes
Gutenberg-Universit\"at, D-55099 Mainz, Germany}
\author{J.~Gegelia}
\affiliation{Institut f\"ur Kernphysik, Johannes
Gutenberg-Universit\"at, D-55099 Mainz, Germany} \affiliation{High
Energy Physics Institute, Tbilisi State University, Tbilisi,
Georgia}
\author{S.~Scherer}
\affiliation{Institut f\"ur Kernphysik, Johannes
Gutenberg-Universit\"at, D-55099 Mainz, Germany}
\begin{abstract}
   We calculate the nucleon form factors $G_A$ and
$G_P$ of the isovector axial-vector current and the pion-nucleon
form factor $G_{\pi N}$ in manifestly Lorentz-invariant
baryon chiral perturbation theory up to and including order ${\cal O}(p^4)$.
   In addition to the standard treatment including the
nucleon and pions, we also consider the axial-vector meson $a_1$
as an explicit degree of freedom.
   This is achieved by using the reformulated infrared
renormalization scheme.
   We find that the inclusion of the axial-vector meson effectively results in
one additional low-energy coupling constant that we determine by a fit to
the data for $G_A$.
   The inclusion of the axial-vector meson results in an improved
description of the experimental data for $G_A$, while the
contribution to $G_P$ is small.
\end{abstract}
\date{\today}
\maketitle

\section{Introduction}

   The electroweak form factors are sets of functions that are used to
parameterize the structure of the nucleon as seen by the electromagnetic and
the weak probes.
   While a wealth of data and theoretical predictions exist for the
electromagnetic form factors (see, e.g.,
\cite{Gao:2003ag,Friedrich:2003iz,Hyde-Wright:2004gh} and references
therein), the nucleon form factors of the isovector axial-vector current,
the axial form factor $G_A(q^2)$ and, in particular, the induced pseudoscalar
form factor $G_P(q^2)$, are not as well-known  (see, e.g.,
\cite{Bernard:2001rs,Gorringe:2002xx} for a review).
   However, there are ongoing efforts to increase our
understanding of these form factors.
   The value of the axial form factor at zero momentum transfer is defined as
the axial-vector coupling constant $g_A$ and is quite precisely
determined from neutron beta decay.
   The $q^2$ dependence of the axial form factor can be obtained
either through neutrino scattering or pion electroproduction
(see \cite{Bernard:2001rs} and references therein).
      The second method makes use of the so-called Adler-Gilman relation
\cite{Adler:1966} which provides a chiral Ward identity establishing a connection between charged
pion electroproduction at threshold and the isovector axial-vector current
evaluated between single-nucleon states (see, e.g.,
\cite{Scherer:1991cy} for more details).
   The induced pseudoscalar form factor $G_P(q^2)$ is even less known than
$G_A(q^2)$.
   It has been investigated in ordinary and radiative muon capture
as well as pion electroproduction.
   Analogous to the axial-vector coupling constant $g_A$, the induced pseudoscalar
coupling constant is defined through $g_P = \frac{m_\mu}{2m_N}G_P(q^2=-0.88
m_\mu^2)$, where $q^2=-0.88\, m_\mu^2$ corresponds to muon capture kinematics
and the additional factor $\frac{m_\mu}{2m_N}$ stems
from a different convention used in muon capture.
   For a comprehensive review on the experimental and theoretical situation
concerning $G_P(q^2)$ see for example \cite{Gorringe:2002xx}.
   A discrepancy between the results in ordinary and radiative
muon capture has recently been addressed in \cite{Clark:2005as}.
   Theoretical approaches to the axial and induced pseudoscalar form factors
include the early current algebra and PCAC calculations
\cite{Adler:1966,Nambu:1997wb,Sato:1967},
various quark model (see, e.g.,
\cite{Tegen:1983gg,Hwang:1984sz,Boffi:2001zb,Merten:2002nz,Ma:2002xu,
Khosonthongkee:2004qm,Silva:2005fa})
and lattice calculations \cite{Liu:1992ab}.
   For a recent discussion on extracting the axial form factor in the
timelike region from $\bar p + n\to \pi^- +\ell^-+\ell^+$
($\ell=e$ or $\mu$)
see \cite{Adamuscin:2006bk}.
   Chiral perturbation theory (ChPT)
\cite{Weinberg:1978kz,Gasser:1984yg,Gasser:1984gg,Gasser:1988rb} is the
low-energy effective theory of the standard model and as such
allows model-independent calculations of nucleon properties (see
\cite{Bernard:1995dp,Scherer:2002tk} for an introduction).
   The axial form factor has been addressed in the framework of
heavy-baryon ChPT
\cite{Bernard:1992ys,Bernard:1993bq,Fearing:1997dp,Bernard:1998gv}.
   In principle, when considering a charged transition there is
a third form factor, the induced
pseudotensorial form factor $G_T(q^2)$.
   As will be explained below, this form factor vanishes when combining
isospin symmetry and charge-conjugation invariance and therefore
is not considered in this work \cite{Weinberg:1958ut}.
   Experimentally the induced pseudotensorial form factor is found
to be small \cite{Wilkinson:2000gx,Minamisono:2001cd}.
   Finally, defining the pion-nucleon form factor in terms
of the pseudoscalar quark density and
using the partially conserved axial-vector current (PCAC)
relation allows one to determine the pion-nucleon form factor, once
the axial and induced pseudoscalar form factors are known.

   In this paper we calculate the axial, the induced
pseudoscalar, and the pion nucleon form factors of the nucleon in manifestly
Lorentz-invariant ChPT up to and including order ${\cal O}(p^4)$.
   The renormalization procedure is performed in the framework of
the infrared renormalization of \cite{Becher:1999he}.
   In its reformulated version \cite{Schindler:2003xv}, this renormalization
scheme allows for the inclusion of further degrees of freedom.
   In the following we will include the $a_1$ axial-vector meson
as an explicit degree of freedom.
   It needs to be pointed out that in a strict chiral expansion up
to order ${\cal O}(p^4)$ the results will not differ from the ones
obtained in the standard framework.
   However, explicitly keeping all terms generated from the considered diagrams
involving the axial-vector meson amounts to a resummation of
higher-order contributions.
   This phenomenological approach has shown an improved
description of the electromagnetic form factors of the nucleon
\cite{Kubis:2000zd,Schindler:2005ke} when the $\rho$, $\omega$,
and $\phi$ mesons are included.

   This paper is organized as follows: In
Sec.~\ref{sec:Def} the definitions and some important properties
of the relevant form factors are given.
   Section~\ref{sec:Lag} contains the effective Lagrangians used in the present
calculation.
   We present and discuss the results for the form factors with and without
the inclusion of the axial-vector meson $a_1$ in
Sec.~\ref{sec:Results}. Section~\ref{sec:Sum} contains a short
summary.

\section{\label{sec:Def}Definition and properties of the isovector axial-vector
current}
   In QCD, the three components of the isovector axial-vector current
are defined as
\begin{equation}
A^{\mu,a}(x)\equiv \bar{q}(x)\gamma^\mu\gamma_5
\frac{\tau^a}{2}q(x), \quad
q=\left(\begin{array}{c}u\\d\end{array}\right),\quad  a=1,2,3.
\end{equation}
    The operators $A^{\mu,a}(x)$ satisfy the following properties
relevant for the subsequent discussion:
\begin{enumerate}
\item Hermiticity:
\begin{equation}\label{Aherm}
    A^{\mu,a\dagger}(x) = A^{\mu,a}(x).
\end{equation}
\item Equal-time commutation relations with
the vector charges:
\begin{equation}\label{AIsospin}
    [Q^a_V(t),A^{\mu,b}(t,\vec x)]=i\epsilon^{abc}A^{\mu,c}(t,\vec x).
\end{equation}
\item Transformation behavior under parity:
\begin{equation}\label{Aparity}
    A^{\mu,a}(x) \stackrel{\cal P}{\mapsto} -A_\mu^a(\tilde x),\quad
\tilde x^\mu=x_\mu.
\end{equation}
\item Transformation behavior under charge conjugation:
\begin{eqnarray}
   A^{\mu,a}(x)&\stackrel{{\cal C}}{\mapsto}&A^{\mu,a}(x),
\quad a=1,3, \nonumber \\
   A^{\mu,2}(x)&\stackrel{{\cal C}}{\mapsto}&-A^{\mu,2}(x).
\end{eqnarray}
\item Partially conserved axial-vector current (PCAC) relation:
\begin{equation}
\label{pcac}
\partial_\mu A^{\mu,a}=i\bar{q}\gamma_5\{\frac{\tau^a}{2},{\cal M}\}q,
\end{equation}
where ${\cal M}=\mbox{diag}(m_u,m_d)$ is the quark mass matrix.
\end{enumerate}
   Assuming isospin symmetry, $m_\mu=m_d=\hat m$, the most general
parametrization of the isovector axial-vector current
evaluated between one-nucleon states in terms of axial-vector covariants
is given by
\begin{equation}\label{FFDef}
\langle N(p')| A^{\mu,a}(0) |N(p) \rangle = \bar{u}(p')
\left[\gamma^\mu\gamma_5 G_A(q^2)
+\frac{q^\mu}{2m_N}\gamma_5 G_P(q^2)
\right]
\frac{\tau^a}{2}u(p),
\end{equation}
where $q_\mu=p'_\mu-p_\mu$ and $m_N$ denotes the nucleon mass.
   $G_A(q^2)$ is called the axial form factor and
$G_P(q^2)$ is the induced pseudoscalar form factor.
   From the Hermiticity of Eq.~(\ref{Aherm}), we find that
$G_A$ and $G_P$ are real for space-like momenta ($q^2\leq 0$).
  In the case of perfect isospin symmetry
the strong interactions are invariant under ${\cal G}$ conjugation, which is a
combination of charge conjugation $\cal C$ and a rotation by $\pi$ about the 2
axis in isospin space (charge symmetry operation),
\begin{equation}
{\cal G}={\cal C}\exp(i\pi Q_V^2).
\end{equation}
   The presence of a third so-called second-class structure
\cite{Weinberg:1958ut} of the type $i\sigma^{\mu\nu}q_\nu\gamma_5$ in the
charged transition would indicate a violation of $\cal G$ conjugation.
  As there seems to be no clear empirical evidence for such a
contribution \cite{Wilkinson:2000gx,Minamisono:2001cd}
we will omit it henceforth.

   Similarly, the nucleon matrix element of the pseudoscalar density
$P^a(x)=i\bar q(x) \gamma_5 \tau^a q(x)$
can be parameterized as
\begin{equation}
\label{GpiN}
\hat m \langle N(p')| P^a (0) | N(p) \rangle =
         \frac{M_\pi^2 F_\pi}{M_\pi^2 - q^2}
         G_{\pi N}(q^2)i\bar{u}(p') \gamma_5 \tau^a u(p),
\end{equation}
   where $M_\pi$ is the pion mass and $F_\pi$ the pion decay constant.
   Equation (\ref{GpiN}) {\em defines} the form factor
$G_{\pi N}(q^2)$ in terms of the QCD operator $\hat m  P^a(x)$.
   The operator $\hat m P^a(x)/(M_\pi^2 F_\pi)$ serves as
an interpolating pion field and thus $G_{\pi N}(q^2)$ is also referred to as
the pion-nucleon form factor for this specific choice of the interpolating
pion field \cite{Bernard:1995dp}.
   The pion-nucleon coupling constant $g_{\pi N}$ is defined through
$G_{\pi N}(q^2)$ evaluated at $q^2=M_\pi^2$.
   As a result of the PCAC relation, Eq.\ (\ref{pcac}), the three form
factors $G_A$, $G_P$, and $G_{\pi N}$ are related by
\begin{equation}
\label{ff_relation}
   2m_N G_A(q^2) + \frac{q^2}{2m_N} G_P(q^2) =
       2\frac{M_\pi^2 F_\pi}{M_\pi^2 - q^2} G_{\pi N}(q^2).
\end{equation}

\section{\label{sec:Lag}Effective Lagrangian and power counting}

   The calculation of the isovector axial-vector current form factors
of the nucleon requires both the purely mesonic as well as the one-nucleon
part of the chiral effective Lagrangian up to order ${\cal O}(p^4)$,
\begin{equation}\label{LagBasic}
{\cal L}_{\rm eff} = {\cal L}_{2} + {\cal L}_{4} + {\cal L}^{(1)}_{{\pi}N} +
   {\cal L}^{(2)}_{{\pi}N} + {\cal L}^{(3)}_{{\pi}N} + {\cal L}^{(4)}_{{\pi}N}
       + \cdots.
\end{equation}
   Here, $p$ collectively stands for a ``small'' quantity such as the
pion mass, a small external four-momentum of the pion or of an
external source, and an external three-momentum of the nucleon.

   The pion fields are contained in the $2\times 2$ matrix $U$,
\begin{eqnarray}\label{U}
   U(x) &=& u^2(x) = \exp\left(\frac{i \Phi(x)}{F}\right), \\
   \Phi &=& \vec{\tau}\cdot \vec{\phi} = \left(
         \begin{array}{cc}
            \pi^0 & \sqrt{2}\pi^+ \\ \sqrt{2}\pi^- & -\pi^0
         \end{array}\right),
\end{eqnarray}
   and the purely mesonic Lagrangian at order ${\cal O}(p^2)$ is given by
\cite{Gasser:1984yg}
\begin{equation}\label{L2}
    {\cal L}_2=\frac{F^2}{4}\mbox{Tr}\left[D_\mu U (D^\mu
U)^\dagger \right] + \frac{F^2}{4}\mbox{Tr}\left[\chi U^\dagger
+ U \chi^\dagger\right].
\end{equation}
   The covariant derivative $D_\mu U$ with a coupling to an external
axial-vector field $a_\mu=\tau^a a^a_\mu/2$ only is given by
\begin{displaymath}\label{covDer}
    D_\mu U = \partial_\mu U -i a_\mu U -i U a_\mu\,,
\end{displaymath}
while $\chi$ is defined as
\begin{displaymath}\label{chi}
   \chi=2B(s+ip),
\end{displaymath}
with $s$ and $p$ the scalar and pseudoscalar external sources,
respectively.
   $F$ denotes the pion decay constant in the chiral limit,
$F_\pi = F[1+{\cal O}(\hat{m})] = 92.42(26)$ MeV  \cite{Yao:2006}.
   We work in the isospin-symmetric limit $m_u=m_d=\hat m$, and
the lowest-order expression for the squared pion mass is
$M^2=2 B\hat m$, where
$B$ is related to the quark condensate
$\langle\bar{q}q\rangle_0$ in the chiral limit
\cite{Gasser:1984yg,Colangelo:2001sp},
$\langle\bar u u\rangle_0=\langle\bar d d\rangle_0=-F^2 B$.

   For the mesonic Lagrangian at order ${\cal O}(p^4)$ we only list
the term that contributes to our calculation,
\begin{equation}\label{L4}
   {\cal L}_4=\cdots + \frac{l_4}{8}\mbox{Tr}\left[D_\mu U (D^\mu
U)^\dagger \right]\mbox{Tr}\left[\chi U^\dagger + U
\chi^\dagger\right] + \cdots\,.
\end{equation}
   The complete list for the $SU(2)$ case can be found in \cite{Gasser:1988rb}.

   The lowest-order pion-nucleon Lagrangian is given by
\cite{Gasser:1988rb}
\begin{equation}\label{LagpiN1}
  {\cal L}^{(1)}_{\pi N}=\bar{\Psi}\left(i D\hspace{-.57em}/\hspace{.1em} -
m
+ \frac{\texttt{g}_A}{2}\,
\gamma^\mu\gamma_5 u_\mu \right) \Psi,
\end{equation}
with $m$ the nucleon mass and
$\texttt{g}_A$ the axial-vector coupling constant both
evaluated in the chiral limit.

   For the nucleonic Lagrangians of higher orders we only display those terms
that contribute to our calculations.
   A complete list of terms at orders ${\cal O}(p^2)$ and ${\cal O}(p^3)$ can
be found in \cite{Gasser:1988rb,Fettes:2000gb}.
   At second order the Lagrangian reads
\begin{eqnarray}\label{LagpiN2}
   {\cal L}_{{\pi}N}^{(2)} &=& c_1 \mbox{Tr}(\chi_{+})\bar\Psi\Psi
      - \frac{c_2}{4m^2}\left[ \bar{\Psi}\mbox{Tr}\left(u_\mu
      u_\nu\right)D^\mu D^\nu \Psi + \mbox{h.c.} \right]
  + \frac{c_3}{2}\,\bar{\Psi}\mbox{Tr}\left(u_\mu u^\mu\right)\Psi\nonumber\\
&&-\frac{c_4}{4}\bar\Psi\gamma^\mu\gamma^\nu [u_\mu ,u_\nu ]\Psi
+\cdots ,
\end{eqnarray}
while at order ${\cal O}(p^3)$ we need
\begin{equation}\label{LagpiN3}
    {\cal L}_{{\pi}N}^{(3)} =
    \frac{d_{16}}{2}\bar{\Psi}\gamma^\mu\gamma_5\mbox{Tr}(\chi_+)u_\mu\Psi
    +\frac{d_{22}}{2}\bar{\Psi}\gamma^\mu\gamma_5
    \left[D_\nu,F_{\mu\nu}^-\right]\Psi
    +\cdots\,.
\end{equation}
   There are no contributions from ${\cal L}^{(4)}_{{\pi}N}$ in our calculation.
   The Lagrangians contain the building blocks
\begin{eqnarray*}\label{Buildblocks}
   D_\mu\Psi &=&\left(\partial_\mu +\Gamma_\mu\right)\Psi,\\
   \Gamma_\mu &=&
     \frac{1}{2}\,\left[ u^\dagger \left( \partial_\mu -ia_\mu \right)u
     + u \left( \partial_\mu -ia_\mu \right)u^\dagger \right],\\
   u_\mu &=&  i\left[ u^\dagger \partial_\mu u -u \partial_\mu
     u^\dagger-i (u^\dagger  a_\mu u + u a_\mu u^\dagger)\right],\\
   \chi_+&=&u^\dagger\chi u^\dagger+u\chi^\dagger u,\\
   F^-_{\mu\nu} &=& u^\dagger(\partial_\mu a_\nu-\partial_\nu a_\mu
-i[a_\mu,a_\nu])u
+u(\partial_\mu a_\nu-\partial_\nu a_\mu
+i[a_\mu,a_\nu])u^\dagger,
\end{eqnarray*}
where we only display the external axial-vector source $a_\mu$.

   In order to include axial-vector mesons as explicit degrees of freedom we
consider the vector-field formulation of \cite{Ecker:yg} in which the
$a_1(1260)$ meson is represented by $A_\mu=A_\mu^a \tau^a$.
   The advantage of this formulation is that the coupling of the axial-vector
mesons to pions and external sources is at least of order ${\cal O}(p^3)$.
   A complete list of possible couplings at this order can be found in
\cite{Ecker:yg}.
   The calculation of the contributions to the isovector axial-vector
form factors only requires the term
\begin{equation}\label{LagAVMmeson}
    {\cal L}_{{\pi}A}^{(3)} = \frac{f_A}{4}
\mbox{Tr}(A_{\mu\nu}F^{\mu\nu}_{-}),
\end{equation}
where
\begin{displaymath}
A_{\mu\nu}=\nabla_\mu A_\nu-\nabla_\nu A_\mu
\end{displaymath}
with
\begin{displaymath}
\nabla_\mu A_\nu=\partial_\mu A_\nu+[\Gamma_\mu,A_\nu].
\end{displaymath}
   The coupling of the axial-vector meson to the nucleon starts at
order ${\cal O}(p^0)$.
   The corresponding Lagrangian reads
\begin{equation}\label{LagAVMNuc}
   {\cal L}_{NA}^{(0)} = \frac{g_{a_1}}{2} \bar{\Psi} \gamma^{\mu}
   \gamma_5 A_\mu \Psi.
\end{equation}
   A calculation up to order ${\cal O}(p^4)$ would in principle
also require the Lagrangian of order ${\cal O}(p)$.
   However, there is no term at this order that is allowed by the symmetries.

   In addition to the usual counting rules for pions and nucleons
(see, e.g., \cite{Scherer:2002tk}), we count the axial-vector
meson propagator as order ${\cal O}(p^0)$, vertices from ${\cal
L}_{{\pi}A}^{(3)}$ as order ${\cal O}(p^3)$ and vertices from
${\cal L}_{AN}^{(0)}$ as order ${\cal O}(p^0)$, respectively
\cite{Fuchs:2003sh}.

\section{\label{sec:Results}Results and Discussion}

\subsection{\label{sec:without}Results without axial-vector mesons}
\begin{figure}[ht]
\epsfig{file=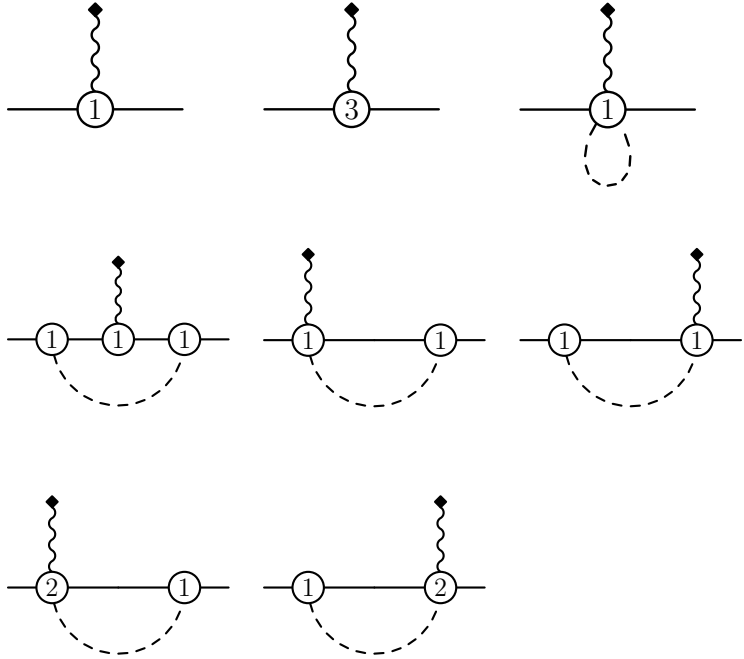,width=0.6\textwidth}\caption{\label{Irreduc}
One-particle-irreducible diagrams contributing to the nucleon matrix
element of the isovector axial-vector current.}
\end{figure}

   The axial form factor $G_A(q^2)$ only receives contributions from
the one-particle-irreducible diagrams of Fig.~\ref{Irreduc}.
   The unrenormalized result reads
\begin{eqnarray}\label{GA}
  {G_A}_0(q^2) &=&\texttt{g}_A +4 M^2 d_{16} -d_{22}q^2
-\frac{\texttt{g}_A}{F^2}I_{\pi}
    +2\frac{\texttt{g}_A}{F^2}M^2 I_{{\pi}N}(m_N^2)
\nonumber \\
  && +8\frac{\texttt{g}_A}{F^2}m_N
\left\{c_4\left[M^2I_{{\pi}N}(m_N^2)-I_{{\pi}N}^{(00)}(m_N^2)\right]
    -c_3 I_{{\pi}N}^{(00)}(m_N^2) \right\}\nonumber \\
  && -\frac{\texttt{g}_A^3}{4F^2}
\left[ I_{\pi}-4m_N^2 I_{{\pi}N}^{(p)}(m_N^2)
+4m_N^2(n-2)I_{{\pi}NN}^{(00)}(q^2)\right.\nonumber\\
&&\left.
    +16m_N^4 I_{{\pi}NN}^{(PP)}(q^2) +4m_N^2 t I_{{\pi}NN}^{(qq)}(q^2)
    \right].
\end{eqnarray}
   The definition of the integrals can be found in the appendix.
   To renormalize the expression for $G_A(q^2)$ we
multiply Eq.~(\ref{GA}) by the nucleon wave function
renormalization constant $Z$ \cite{Becher:1999he},
\begin{equation}\label{ZNuc}
    Z = 1-\frac{9 \texttt{g}_A^2 M^2}{32\pi^2 F^2}
\left[\frac{1}{3}+\ln\left(\frac{M}{m}\right)\right]+
    \frac{9 \texttt{g}_A^2 M^3}{64\pi F^2 m},
\end{equation}
and replace the integrals with their infrared singular parts.

   The axial-vector coupling constant $g_A$ is defined as $g_A=G_A(q^2=0)
=1.2695(29)$ \cite{Yao:2006} and we obtain for its quark-mass expansion
\begin{equation}
\label{gAexpand}
g_A=\texttt{g}_A+g_A^{(1)}M^2+g_A^{(2)}M^2\ln\left(\frac{M}{m}\right)
+g_A^{(3)} M^3 +{\cal O}(M^4),
\end{equation}
with
\begin{eqnarray}
  g_A^{(1)} &=& 4d_{16}-\frac{\texttt{g}_A^3}{16\pi^2F^2}\,, \nonumber\\
  g_A^{(2)} &=&
-\frac{\texttt{g}_A}{8\pi^2F^2}(1+2\texttt{g}_A^2)\,, \nonumber\\
  g_A^{(3)} &=& \frac{\texttt{g}_A}{8\pi F^2 m}(1+\texttt{g}_A^2)
-\frac{\texttt{g}_A}{6\pi F^2}(c_3-2c_4),
\end{eqnarray}
   where all coefficients are understood as IR renormalized parameters.
    These results agree with the chiral coefficients obtained in
HBChPT \cite{Kambor:1998pi,Bernard:2006te} as well as the IR calculation
of \cite{Ando:2006xy}.
   It is worth noting that an agreement for the analytic term
$g_A^{(1)}$ cannot be expected in general.
   For example, when expressed in terms of the renormalized couplings
of the extended on-mass-shell (EOMS) renormalization scheme of
\cite{Fuchs:2003qc}, the $g_A^{(1)}$ coefficient is given by
\cite{Ando:2006xy}
\begin{displaymath}
4 d_{16}^{EOMS}-\frac{\texttt{g}_A}{16\pi^2 F^2}(2+3\texttt{g}_A^2)
+\frac{c_1 \texttt{g}_A m}{4\pi^2 F^2}(4-\texttt{g}_A^2).
\end{displaymath}
   Such a difference is not a surprise, because the use of different
renormalization schemes is compensated by different values of the
renormalized parameters.
   For a similar discussion regarding the chiral expansion of the
nucleon mass, see \cite{Fuchs:2003qc}.

   The axial form factor can be written as
\begin{equation}\label{GAexpansion}
    G_A(q^2)=g_A+\frac{1}{6}\,g_A\,\langle r^2_A\rangle\, q^2 +
    \frac{\texttt{g}_A^3}{4F^2}H(q^2),
\end{equation}
where $\langle r^2_A\rangle$ is the axial mean-square radius
and $H(q^2)$ contains loop contributions and satisfies $H(0)=H'(0)=0$.
   The low-energy coupling constants (LECs) $d_{16}$ and $d_{22}$ are thus
absorbed in the axial-vector coupling constant $g_A$ and the axial
mean-square radius $\langle r^2_A\rangle$.
  The numerical contribution of $H(q^2)$ is negligible which can be
understood by expanding $H$ in a Taylor series in $q^2$.
   Such an expansion generates powers of $q^2/m^2$ where the individual
coefficients have a chiral expansion similar to Eq.~(\ref{gAexpand}).

   For the analysis of experimental data, $G_A(q^2)$ is conventionally
parameterized using a dipole form as
\begin{equation}\label{GAPara}
    G_A(q^2)=\frac{g_A}{(1-\frac{q^2}{M^2_A})^2},
\end{equation}
where the so-called axial mass $M_A$ is related to the axial root-mean-square
radius by $\langle r^2_A\rangle^\frac{1}{2}=2\sqrt{3}/M_A$.
    The global average for the axial mass extracted from neutrino scattering
experiments given in \cite{Liesenfeld:1999mv} is
\begin{equation}\label{MAv1}
    M_A = (1.026 \pm 0.021)\,\mbox{GeV},
\end{equation}
whereas a recent analysis \cite{Budd:2003wb} taking account of updated
expressions for the vector form factors finds a slightly smaller value
\begin{equation}\label{MAv2}
    M_A = (1.001 \pm 0.020)\,\mbox{GeV}.
\end{equation}
   On the other hand, smaller values of $(0.95\pm 0.03)$ GeV and
$(0.96\pm 0.03)$ GeV have been obtained in \cite{Kuzmin:2006dh} as world
averages from quasielastic scattering and $(1.12\pm 0.03)$ GeV
from single pion neutrinoproduction.
   Finally, the most recent result extracted
from quasielastic $\nu_\mu n\to \mu^- p$ in oxygen nuclei reported by
the K2K Collaboration, $M_A=(1.20\pm 0.12)$ GeV, is considerably
larger \cite{Gran:2006jn}.

   The extraction of the axial mean-square radius from
charged pion electroproduction at threshold is motivated by
the current algebra results and the PCAC hypothesis.
   The most recent result for the reaction $p(e,e'\pi^+)n$ has been obtained
at MAMI at an invariant mass
of $W=1125$ MeV (corresponding to a pion center-of-mass momentum of
$|\vec{q}^\ast|=112$ MeV) and photon four-momentum transfers of $-k^2=0.117$,
$0.195$ and 0.273 GeV$^2$ \cite{Liesenfeld:1999mv}.
   Using an effective-Lagrangian model
an axial mass of
\begin{displaymath}
\bar{M}_A=(1.077\pm 0.039)\,\mbox{GeV}
\end{displaymath}
was extracted, where the bar is used to distinguish the result from the
neutrino scattering value.
    In the meantime, the experiment has been repeated including an
additional value of $-k^2=0.058$ GeV$^2$ \cite{Baumann:2004} and is currently
being analyzed.
   The global average from several pion electroproduction experiments
is given by \cite{Bernard:2001rs}
\begin{equation}\label{MApi}
    \bar{M}_A=(1.068\pm 0.017)\,\mbox{GeV}.
\end{equation}
   It can be seen that the values of Eqs.~(\ref{MAv1}) and
(\ref{MAv2}) for the neutrino scattering experiments are smaller
than that of Eq.~(\ref{MApi}) for the pion electroproduction
experiments.
   The discrepancy was explained in heavy baryon chiral
perturbation theory \cite{Bernard:1992ys}.
   It was shown that at order ${\cal O}(p^3)$ pion
loop contributions modify the $k^2$ dependence of the electric dipole amplitude
from which $\bar{M}_A$ is extracted.
   These contributions result in a change of
\begin{equation}\label{deltaMA}
    \Delta M_A = 0.056 \,\mbox{GeV},
\end{equation}
bringing the neutrino scattering and pion electroproduction
results for the axial mass into agreement.

\begin{figure}[ht]
\epsfig{file=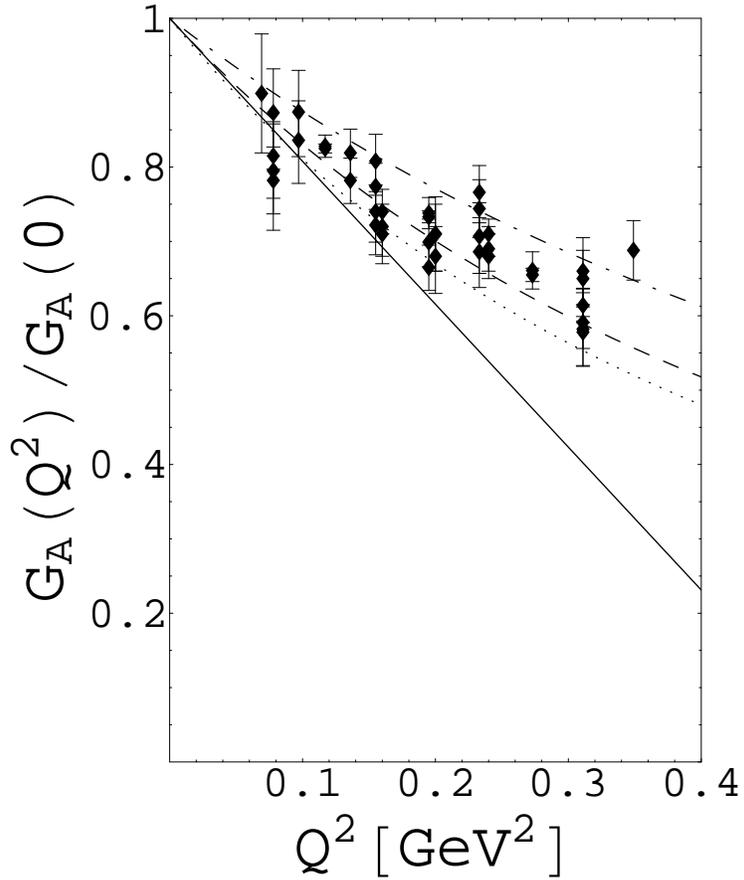,width=0.6\textwidth}\caption{\label{GAwithout}
The axial form factor $G_A$ in manifestly Lorentz-invariant ChPT
at ${\cal O}(p^4)$. Full line: result in infrared renormalization with
parameters fitted to reproduce the axial mean-square radius corresponding
to the dipole parametrization with $M_A=1.026$ GeV (dashed line).
  The dotted and dashed-dotted lines refer to dipole parameterizations
with $M_A=0.95$ GeV and $M_A=1.20$ GeV, respectively.
   The experimental values are
taken from \cite{Bernard:2001rs}.}
\end{figure}

   Using the convention $Q^2=-q^2$ the result for the axial form factor
$G_A(q^2)$ in the momentum
transfer region $0\,\mbox{GeV}^2\leq Q^2 \leq 0.4\,\mbox{GeV}^2$
is shown in Fig.~\ref{GAwithout}.
   The parameters have been determined such as to reproduce the axial
mean-square radius corresponding to the dipole parameterization with
$M_A=1.026$ GeV (dashed line).
  The dotted and dashed-dotted lines refer to dipole parameterizations
with $M_A=0.95$ GeV and $M_A=1.20$ GeV, respectively.
   As anticipated, the loop contributions from $H(q^2)$ are small
and the result does not produce enough curvature to describe the data
for momentum transfers $Q^2 \ge 0.1\, \mbox{GeV}^2$.
   The situation is reminiscent of the electromagnetic case
\cite{Kubis:2000zd,Fuchs:2003ir} where ChPT at ${\cal O}(p^4)$
also fails to describe the form factors beyond $Q^2 \ge 0.1\,
\mbox{GeV}^2$.

\begin{figure}[ht]
\epsfig{file=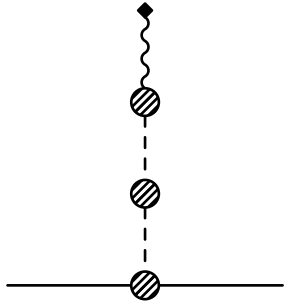,width=0.23\textwidth}\caption{\label{PiPoleDia}
Pion pole graph of the isovector axial-vector current.}
\end{figure}

   The one-particle-irreducible diagrams of Fig.~\ref{Irreduc} also contribute
to the induced pseudoscalar form factor $G_P(q^2)$,
\begin{equation}\label{GPirr}
    G_P^{irr}(q^2) = 4m_N^2 d_{22} + 8m_N^4
    \frac{g_A^3}{F^2}I_{{\pi}NN}^{(qq)}(q^2)\,.
\end{equation}
  Furthermore, $G_P(q^2)$ receives contributions from the pion pole graph of
Fig.~\ref{PiPoleDia}.
   It consists of three building blocks: The coupling of the external axial
source to the pion, the pion propagator, and the ${\pi}N$-vertex,
respectively.
   We consider each part separately.

\begin{figure}[ht]
\epsfig{file=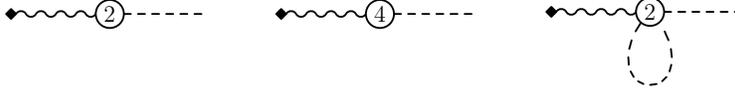,width=0.6\textwidth}\caption{\label{APiDia}
Diagrams contributing to the coupling of the isovector
axial-vector current to a pion up to order ${\cal O}(p^4)$.}
\end{figure}

   The renormalized coupling of the external axial source to a pion
up to order ${\cal O}(p^4)$ is given by
\begin{equation}\label{api}
    \epsilon_A \cdot q F_\pi \delta_{ij},
\end{equation}
where the diagrams in Fig.~\ref{APiDia} have been taken into
account and the renormalized pion decay constant reads
\begin{equation}\label{Frenorm}
    F_\pi = F \left[ 1+\frac{M^2}{F^2}l_4^r
- \frac{M^2}{8\pi^2F^2}\ln\left(\frac{M}{m}\right)+{\cal O}(M^4)
    \right].
\end{equation}
   We have used the pion wave function renormalization constant
\begin{equation}\label{Zpi}
    Z_\pi = 1-\frac{2M^2}{F^2}\left[ l_4^r+\frac{1}{24\pi^2}\,\left(
    R-\ln\left(\frac{M}{m}\right) \right) \right],
\end{equation}
with $l_4^r$ the renormalized coupling of Eq.~(\ref{L4}) and
$R=\frac{2}{n-4}+\gamma_E-1-\ln(4\pi)$.

   The renormalized pion propagator is obtained by simply
replacing the lowest-order pion mass $M$ by the expression for the
physical mass $M_\pi$ up to order ${\cal O}(p^4)$,
\begin{equation}\label{Mphys}
    M^2_\pi = M^2+\Sigma(M^2_\pi) = M^2\left[ 1 +\frac{2M^2}{F^2}\left( l_3^r
      +\frac{1}{32\pi^2}\ln\left(\frac{M}{m}\right) \right)
      \right].
\end{equation}

\begin{figure}[ht]
\epsfig{file=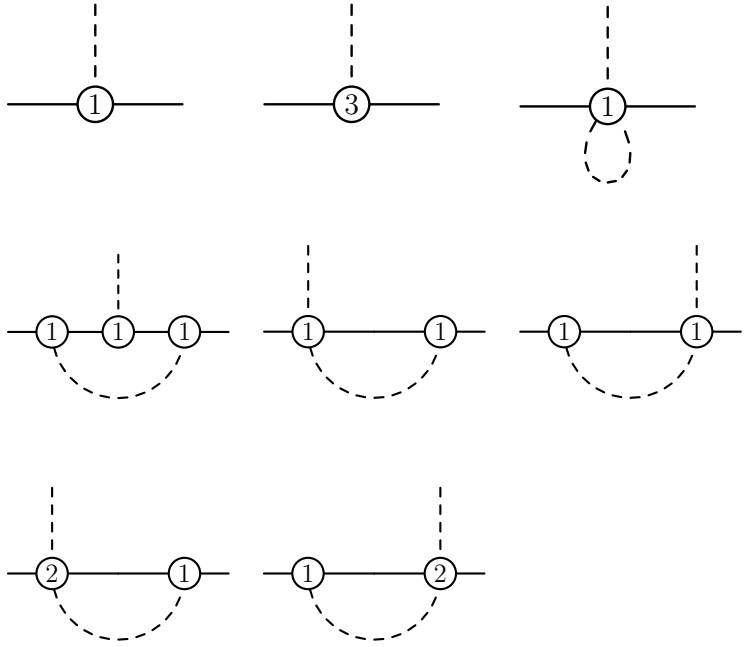,width=0.6\textwidth}\caption{\label{PiNDia}
Diagrams contributing to the pion-nucleon vertex up to order
${\cal O}(p^4)$.}
\end{figure}

   The ${\pi}N$ vertex evaluated between on-mass-shell nucleon states
up to order ${\cal O}(p^4)$ receives contributions from the
diagrams in Fig.~\ref{PiNDia} and the unrenormalized result for a
pion with isospin index $i$ is given by
\begin{eqnarray}\label{PiNVertex}
  \Gamma(q^2)\gamma_5 \tau_i &=& \left( -\frac{\texttt{g}_A}{F}m_N
    +2\frac{M^2}{F}m_N(d_{18}-2d_{16}) +\frac{\texttt{g}_A}{3F^2}m_N I_{\pi}
    -2\frac{\texttt{g}_A}{F^3}M^2 m_N I_{{\pi}N}(m_N^2)\right. \nonumber \\
  && -8\frac{\texttt{g}_A}{F^2}m_N^2
\left\{c_4\left[M^2I_{{\pi}N}(m_N^2)-I_{{\pi}N}^{(00)}(m_N^2)\right]
    -c_3 I_{{\pi}N}^{(00)}(m_N^2) \right\} \nonumber \\
  &&\left. +\frac{\texttt{g}_A^3}{4F^3}m_N\left[ I_{\pi} +4mN^2 I_{NN}(q^2)
+4m_N^2 M^2 I_{{\pi}NN}(q^2)
  \right]\right)\gamma_5 \tau_i\,.
\end{eqnarray}
   To find the renormalized vertex one multiplies with
$Z\sqrt{Z_\pi}$ and replaces the integrals with their infrared
singular parts.
   However, the renormalized result should not be confused with the
pion-nucleon form factor $G_{\pi N}(q^2)$ of Eq.\ (\ref{GpiN}).
   In general, the pion-nucleon vertex depends on the choice of the field
variables in the (effective) Lagrangian.
   In the present case, the pion-nucleon vertex is only an auxiliary
quantity, whereas the ``fundamental'' quantity (entering chiral Ward
identities) is the matrix element of the pseudoscalar density.
      Only at $q^2=M_\pi^2$, we expect the same coupling strength, since
both $\hat m P^a(x)/(M_\pi^2 F_\pi)$ and the field
$\phi_i$ of Eq.\ (\ref{U})
serve as interpolating pion fields.
   After renormalization, we obtain for the pion-nucleon coupling
constant the quark-mass expansion
\begin{equation}
\label{gpiNexpand}
    g_{\pi N}=\texttt{g}_{\pi N}
              + g_{{\pi}N}^{(1)} M^2
              + g_{{\pi}N}^{(2)} M^2 \ln\left(\frac{M}{m}\right)
              + g_{{\pi}N}^{(3)}M^3
              +{\cal O}(M^4)\,,
\end{equation}
with
\begin{eqnarray}
  \texttt{g}_{\pi N}&=&\frac{\texttt{g}_A m}{F}\,,\nonumber\\
  g_{{\pi}N}^{(1)}&=&-\texttt{g}_A\frac{l_4^r  m}{F^3}
                     - 4\texttt{g}_A\frac{c_1}{F}
                     +\frac{2(2d_{16}-d_{18})m}{F}
                     -\texttt{g}_A^3\frac{m}{16\pi^2F^3} \,, \nonumber\\
  g_{{\pi}N}^{(2)} &=&-\texttt{g}_A^3\frac{m}{4\pi^2F^3} \,, \nonumber\\
  g_{{\pi}N}^{(3)} &=&\texttt{g}_A\frac{4+ \texttt{g}_A^2}{32\pi F^3}
                      -\texttt{g}_A\frac{(c_3-2c_4)m}{6\pi F^3}\,,
\end{eqnarray}
   where all coefficients are understood as IR renormalized parameters.
    These results agree with the chiral coefficients obtained in
\cite{Becher:2001hv}.
   In the chiral limit, Eq.\ (\ref{gpiNexpand}) satisfies the
Goldberger-Treiman relation $\texttt{g}_{\pi N}=\texttt{g}_A m/F$ \cite{GT}.
   The numerical violation of the Goldberger-Treiman relation as expressed
in the so-called Goldberger-Treiman discrepancy \cite{Pagels:1969ne},
\begin{equation}
\label{GTdiscrepancy}
\Delta=1-\frac{m_N g_A}{F_\pi g_{\pi N}},
\end{equation}
is at the percent level, $\Delta=(2.44^{+0.89}_{-0.51})$ \% for
$m_N=(m_p+m_n)/2=938.92$ MeV, $g_A=1.2695(29)$, $F_\pi=92.42(26)$
MeV, and $g_{\pi N}=13.21^{+0.11}_{-0.05}$ \cite{Schroder:rc}.
   Using different values for the pion-nucleon coupling constant such as
$g_{\pi N}= 13.0\pm 0.1$ \cite{Stoks:1992ja}, $g_{\pi N}= 13.3\pm
0.1$ \cite{Ericson:2000md}, and $g_{\pi N}= 13.15\pm 0.01$
\cite{Arndt:2006bf}
 results in the GT discrepancies $\Delta=(0.79\pm 0.84)$ \%,
$\Delta=(3.03\pm 0.81)$ \%, and  $\Delta=(1.922\pm 0.363)$ \%,
respectively.
   The chiral expansions of $g_A$ etc.~may be used to relate the
parameter $d_{18}$ to $\Delta$ \cite{Becher:2001hv},
\begin{equation}
\label{Delta_d_18}
\Delta=-\frac{2 d_{18} M^2}{\texttt{g}_A}+{\cal O}(M^4).
\end{equation}
   Note that $\Delta$ of Eq.\ (\ref{GTdiscrepancy}) and
$\Delta_{GT}$ of \cite{Becher:2001hv,Schroder:rc} are related
by $\Delta_{GT}=\Delta/(1-\Delta)$.
   In particular, the leading order of the quark mass expansions of
$\Delta$ and $\Delta_{GT}$ is the same.

   The induced pseudoscalar form factor $G_P(q^2)$ is obtained by
combining Eqs.~(\ref{GPirr}), (\ref{Frenorm}), (\ref{Mphys}) and
the renormalized expression for Eq.~(\ref{PiNVertex}).
   With the help of Eqs.\ (\ref{GTdiscrepancy}) and (\ref{Delta_d_18})
it can  entirely be written in terms of known physical
quantities as \cite{Bernard:1994wn}
\begin{equation}\label{GPwoResult}
G_P(q^2)=-4\frac{m_N F_\pi g_{{\pi}N}}{q^2-M^2_\pi}-\frac{2}{3}m_N^2 g_A
\langle r_A^2\rangle + {\cal O}(p^2).
\end{equation}
   The $1/(q^2-M_\pi^2)$ behavior of $G_P$ is not in conflict with
the book-keeping of a calculation
at chiral order ${\cal O}(p^4)$, because
the external axial-vector field $a_\mu$ counts
as ${\cal O}(p)$, and the definition of the matrix element contains a
momentum $(p'-p)^\mu$ and the Dirac matrix $\gamma_5$
so that the combined order of all ingredients in the matrix element ranges
from ${\cal O}(p)$ to ${\cal O}(p^4)$.
   The terms that have been neglected in the form factor $G_P$ are of
order $M^2$, $q^2/m^2$ and higher.

   Using the above values for $m_N$, $g_A$, $F_\pi$ as well as
$g_{{\pi}N}=13.21^{+0.11}_{-0.05}$, $M_A=(1.026\pm 0.021)$ GeV,
$M=M_{\pi^+}=139.57$ MeV and $m_\mu=105.66$ MeV \cite{Yao:2006} we
obtain for the induced pseudoscalar coupling
\begin{equation}\label{gP}
    g_P = 8.29^{+0.24}_{-0.13}\pm 0.52,
\end{equation}
which is in agreement with the heavy-baryon results $8.44\pm 0.23$
\cite{Bernard:1994wn} and $8.21\pm 0.09$ \cite{Fearing:1997dp}, once the
differences in the coupling constants used are taken in consideration.
   The first error given in Eq.~(\ref{gP}) stems only from the empirical
uncertainties in the quantities of Eq.~(\ref{GPwoResult}).
   As an attempt to estimate the error originating in the truncation
of the chiral expansion in the baryonic sector we assign a relative error
of $0.5^k$, where $k$ denotes the diffence between the order that has
been neglected and the leading order at which a nonvanishing result appears.
   Such a (conservative) error is motivated by, e.~g., the analysis
of the individual terms of Eq.~(\ref{gAexpand}) as well as the
determination of the LECs $c_i$ at ${\cal O}(p^2)$ and to one-loop accuracy
${\cal O}(p^3)$ in the heavy-baryon framework \cite{Bernard:1997gq}.
   For $g_P$ we have thus added a truncation error of 0.52.

\begin{figure}[ht]
\epsfig{file=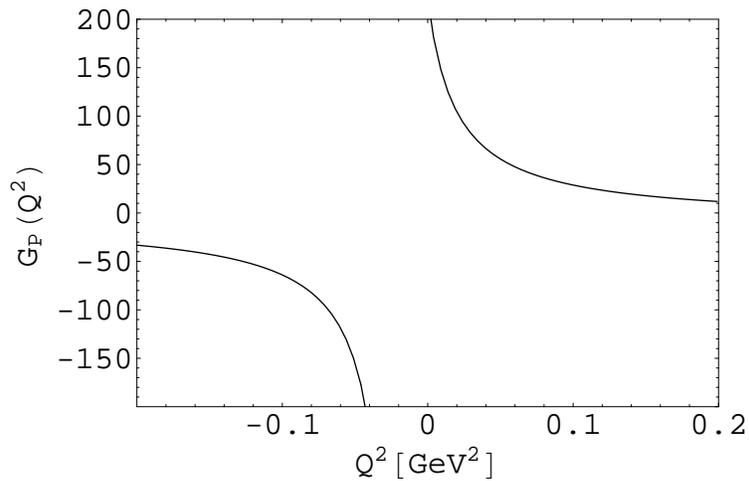,width=0.6\textwidth}\caption{\label{GPwithout}
The induced pseudoscalar form factor $G_P$ in manifestly
Lorentz-invariant ChPT at ${\cal O}(p^4)$.}
\end{figure}

   Figure \ref{GPwithout} shows our result for $G_P(q^2)$ in the
momentum transfer region $-0.2\,\mbox{GeV}^2\leq Q^2 \leq
0.2\,\mbox{GeV}^2$.
   One can clearly see the dominant pion pole contribution at
$q^2\approx M^2_\pi$
which is also supported by the experimental results of
\cite{Choi:1993vt}.

   Using Eq.\ (\ref{ff_relation}) allows one to also determine the
pion-nucleon form factor $G_{\pi N}(q^2)$ in terms of the results for
$G_A(q^2)$ and $G_P(q^2)$.
   When expressed in terms of physical quantities, it has the particularly
simple form
\begin{equation}
\label{GpiNresult}
G_{\pi N}(q^2)=\frac{m_N g_A}{F_\pi}+g_{\pi N}\Delta \frac{q^2}{M_\pi^2}
+{\cal O}(p^4).
\end{equation}
   We have explicitly verified that the results agree with a direct
calculation of $G_{\pi N}(q^2)$ in terms of a coupling to an external
pseudoscalar source.
   Observe that, with our definition in terms of QCD bilinears, the
pion-nucleon form factor is, in general, {\em not} proportional
to the axial form factor.
   The relation $G_{\pi N}(q^2)=m_N G_A(q^2)/F_\pi$ which is sometimes
used in PCAC applications implies a pion-pole dominance for
$G_P(q^2)$ of the form $G_P(q^2)=4m_N^2 G_A(q^2)/(M_\pi^2-q^2)$.
   However, as can be seen from Eq.\
(\ref{GpiNresult}), there are deviations at
${\cal O}(p^2)$ from such a complete pion-pole dominance assumption.

   The difference between $G_{\pi N}(q^2=M_\pi^2)$ and $G_{\pi N}(q^2=0)$
is entirely given in terms of the GT discrepancy \cite{Bernard:1995dp}
\begin{equation}\label{GpNDiff}
   G_{\pi N}(M_\pi^2)-G_{\pi N}(0)=g_{\pi N}\Delta.
\end{equation}
   Parameterizing the form factor in terms of a monopole,
\begin{equation}\label{Monopole}
    G_{\pi N}^{{\rm mono}}(q^2) = g_{\pi
    N}\frac{\Lambda^2-M^2}{\Lambda^2-q^2}\,,
\end{equation}
Eq.~(\ref{GpNDiff}) translates into a mass parameter $\Lambda=894$
MeV for $\Delta=2.44$ \%.

\subsection{\label{sec:with}Inclusion of the axial-vector meson $a_1(1260)$}

\begin{figure}[ht]
\epsfig{file=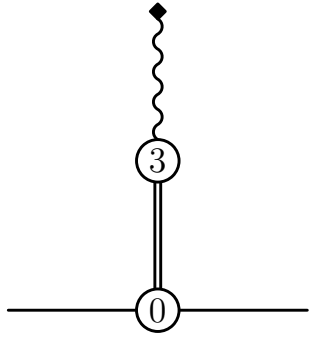,width=0.25\textwidth}\caption{\label{AVMDia}
Diagram containing axial-vector meson (double line) contributing
to the form factors $G_A$ and $G_P$.}
\end{figure}

\begin{figure}[ht]
\epsfig{file=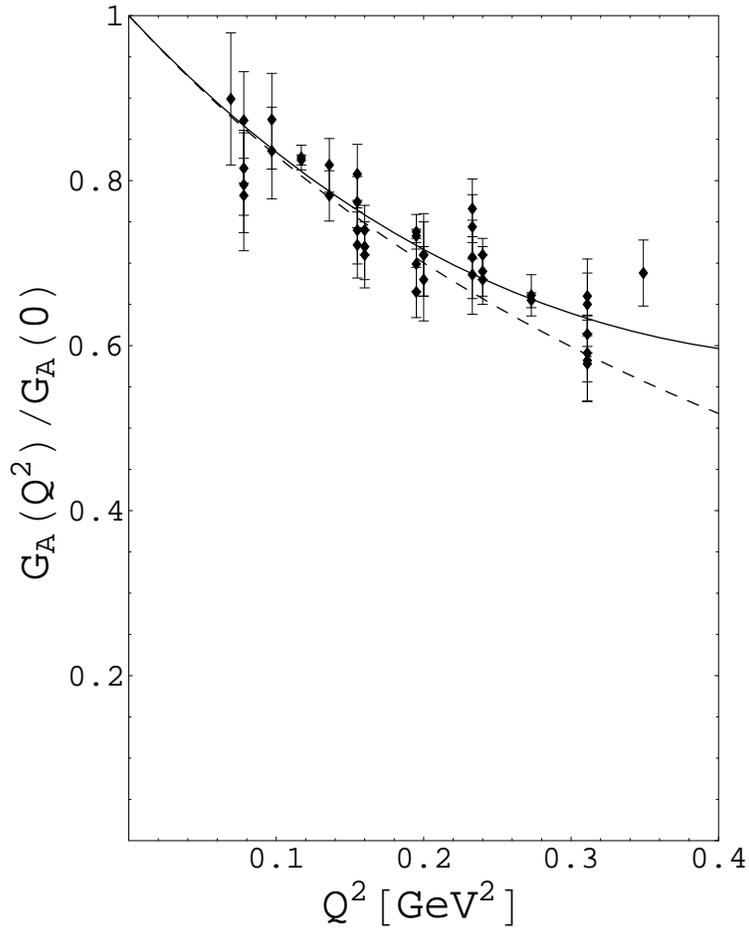,width=0.6\textwidth}\caption{\label{GAwith}
The axial form factor $G_A$ in manifestly Lorentz-invariant ChPT
at ${\cal O}(p^4)$ including the axial-vector meson $a_1$
explicitly. Full line: result in infrared renormalization, dashed
line: dipole parametrization. The experimental values are taken
from \cite{Bernard:2001rs}.}
\end{figure}

   The contributions of the axial-vector meson to the form
factors $G_A$ and $G_P$ at order ${\cal O}(p^4)$ stem from the diagram in
Fig.~\ref{AVMDia}.
   We do not consider loop diagrams with internal axial-vector meson
lines that do not contain internal pion lines, as these vanish in
the infrared renormalization employed in this work.
   With the Langrangians of Eqs.~(\ref{LagAVMmeson}) and (\ref{LagAVMNuc}) the
axial form factor receives
the contribution
\begin{equation}\label{GAAVM}
    G_A^{AVM}(q^2) = - f_A g_{a_1} \frac{q^2}{q^2-M_{a_1}^2}\,,
\end{equation}
   while the result for the induced pseudoscalar form factor
reads
\begin{equation}\label{GPAVM}
    G_P^{AVM}(t) = 4 m_N^2 f_A g_{a_1} \frac{1}{q^2-M_{a_1}^2}\,.
\end{equation}
   The Lagrangians for the axial-vector meson contain two new LECs, $f_A$ and
$g_{a_1}$, respectively.
   However, we find that they only appear through the combination
$f_A g_{a_1}$, effectively leaving only one unknown LEC.
   Performing a fit to the data of $G_A(q^2)$ in the momentum region
$0\,\mbox{GeV}^2\leq Q^2 \leq 0.4\,\mbox{GeV}^2$ the product of
the coupling constants is determined to be
\begin{equation}\label{ConstantFitted}
    f_A g_{a_1} \approx  8.70.
\end{equation}
   Fig.~\ref{GAwith} shows our fitted result for the axial form factor
$G_A(q^2)$ at order ${\cal O}(p^4)$ in the momentum region
$0\,\mbox{GeV}^2\leq Q^2 \leq 0.4\,\mbox{GeV}^2$ with the $a_1$
meson included as an explicit degree of freedom.
   As was expected from phenomenological considerations, the
description of the data has improved for momentum transfers
$Q^2\gtrsim 0.1\,\mbox{GeV}^2$.
   We would like to stress again that in a strict chiral expansion
up to order ${\cal O}(p^4)$ the results with and without axial
vector mesons do not differ from each other.
   The improved description of the data in the case with the
explicit axial-vector meson is the result of a resummation of
certain higher-order terms.
   While the choice of which additional degree of freedom to
include compared to the standard calculation is completely
phenomenological, once this choice has been made there exists a
systematic framework in which to calculate the corresponding
contributions as well as higher-order corrections.

\begin{figure}[ht]
\epsfig{file=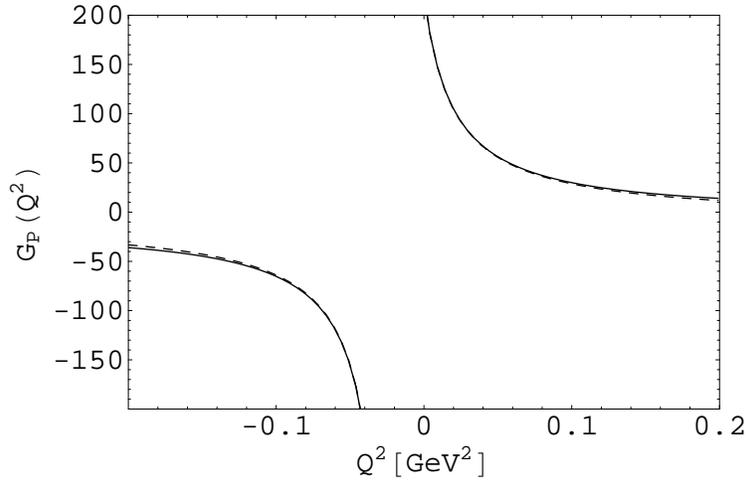,width=0.6\textwidth}\caption{\label{GPwith}
The induced pseudoscalar form factor $G_P$ in manifestly
Lorentz-invariant ChPT at ${\cal O}(p^4)$ including the axial-vector meson
$a_1$ explicitly. Full line: result with axial-vector
meson, dashed line: result without axial-vector meson.}
\end{figure}

   It can be seen from Eq.~(\ref{GAAVM}) that in our formalism the
axial-vector meson does not contribute to the axial-vector
coupling constant $g_A$.
   The pion-nucleon vertex also remains unchanged at the given order,
while the axial mean-square radius receives a contribution.
   The values for the LECs $d_{16}$ and $d_{18}$ therefore do not change,
while $d_{22}$ can be determined from the new expression for the
axial radius using the value of Eq.~(\ref{ConstantFitted}) for the
combination of coupling constants.
   In Fig.~\ref{GPwith} we show the result for $G_P(q^2)$ in the
momentum transfer region $-0.2\,\mbox{GeV}^2\leq Q^2 \leq
0.2\,\mbox{GeV}^2$.
   Also shown for comparison is the result without the explicit
axial-vector meson.
   One sees that the contribution of the $a_1$ to $G_P(q^2)$ for
these momentum transfers is rather small and that $G_P(q^2)$ is
still dominated by the pion pole diagrams.

   The form factors $G_A$ and $G_P$ are related to the
pion-nucleon form factor via Eq.~(\ref{ff_relation}).
   For the contributions of the axial-vector meson we find
\begin{equation}\label{WAAVM}
   2m_N G_A^{AVM}(q^2)+\frac{q^2}{2m_N} G_P^{AVM}(q^2) = 0\,,
\end{equation}
   so that the pion-nucleon form factor is not modified by
the inclusion of the $a_1$ meson.

\section{Summary}\label{sec:Sum}
   We have discussed the nucleon form factors $G_A$ and
$G_P$ of the isovector axial-vector current in manifestly Lorentz-invariant
baryon chiral perturbation theory up to and including order ${\cal O}(p^4)$.
   The main features of the results are similar to the case of
the electromagnetic form factors at the one-loop level.

   As far as the axial form factor is concerned, ChPT can neither predict
the axial-vector coupling constant $g_A$ nor the mean-square axial radius
$\langle r^2_A\rangle$.
   Instead, empirical information on these quantities is used to absorb the
relevant LECs $d_{16}$ and $d_{22}$ in $g_A$ and $\langle r^2_A\rangle$.
   Moreover, the use of a manifestly Lorentz-invariant framework does not
lead to an improved description in comparison with the heavy-baryon
framework, because the re-summed higher-order contributions are negligible.

   The induced pseudoscalar form factor $G_P$ is completely
fixed from ${\cal O}(p^{-2})$ up to and including ${\cal O}(p)$,
once the LEC $d_{18}$ has been expressed in terms of the Goldberger-Treiman
discrepancy.
   Using $g_{\pi N}=13.21$ for the pion-nucleon coupling constant,
we obtain for the induced pseudoscalar coupling
$g_P = 8.29^{+0.24}_{-0.13}\pm 0.52$.
   The first error is due to the error of the empirical quantities entering
the expression for $g_P$ and the second error represents our estimate for the
truncation in the chiral expansion.

   Defining the pion field in terms of the PCAC relation allows one to
introduce a pion-nucleon form factor which is entirely determined in terms
of the axial and induced pseudoscalar form factors.
   Assuming this pion-nucleon form factor to be proportional to the
axial form factor leads to a restriction for $G_P$ which is not supported
by the most general structure of ChPT.

   In addition to the standard treatment including the
nucleon and pions, we have also considered the axial-vector meson $a_1$
as an explicit degree of freedom.
   This was achieved by using the reformulated infrared
renormalization scheme.
   The inclusion of the axial-vector meson effectively results in
one additional low-energy coupling constant which we have determined by a fit
to the data for $G_A$.
   The inclusion of the axial-vector meson results in a considerably improved
description of the experimental data for $G_A$ for values of $Q^2$ up to
about $0.4$ GeV$^2$, while the contribution to $G_P$ is small.

\acknowledgments

   M.R.S.~and S.S.~would like to thank H.W.~Fearing and J.~Gasser for
useful discussions and the TRIUMF theory group for
their hospitality.
   This work was made possible by the financial support from the
Deutsche Forschungsgemeinschaft (SFB 443), the Government of Canada,
and the EU Integrated Infrastructure Initiative Hadron
Physics Project (contract number RII3-CT-2004-506078).

\appendix
\section{\label{sec:IntDef}Definition of loop integrals}
   For the definition of the loop integrals in the expressions
for the form factors we use the notation
\begin{displaymath}
    P^\mu = p_i^\mu + p_f^\mu, \quad q^\mu = p_f^\mu - p_i^\mu.
\end{displaymath}
   Using dimensional regularization \cite{'tHooft:1972fi} the loop integrals
with one or two internal lines are defined as
\begin{eqnarray*}
I_{\pi} &=& i\int\frac{d^nk}{(2\pi)^n} \frac{1}{k^2-M^2+i\epsilon},\\
I_N &=& i\int\frac{d^nk}{(2\pi)^n} \frac{1}{k^2-m^2+i\epsilon},\\
I_{NN}(q^2) &=&
   i\int\frac{d^nk}{(2\pi)^n}
   \frac{1}{[k^2-m^2+i\epsilon][(k+q)^2-m^2+i\epsilon]},\\
I_{{\pi}N}(p^2)&=&
   i\int\frac{d^nk}{(2\pi)^n}
   \frac{1}{[k^2-M^2+i\epsilon][(k+p)^2-m^2+i\epsilon]},\\
p^{\mu}I_{{\pi}N}^{(p)}(p^2) &=&
   i\int\frac{d^nk}{(2\pi)^n}
   \frac{k^{\mu}}{[k^2-M^2+i\epsilon][(k+p)^2-m^2+i\epsilon]},\\
g^{\mu\nu}I_{{\pi}N}^{(00)}(p^2)+p^\mu p^\nu I_{{\pi}N}^{(pp)}(p^2)
   &=& i\int\frac{d^nk}{(2\pi)^n}
\frac{k^{\mu}k^{\nu}}{[k^2-M^2+i\epsilon][(k+p)^2-m^2+i\epsilon]}.
\end{eqnarray*}
   For integrals with three internal lines we assume on-shell
kinematics, $p_f^2=p_i^2=m_N^2$,
\begin{eqnarray*}
I_{{\pi}NN}(q^2) &=& i\int\frac{d^nk}{(2\pi)^n} \frac{1}
   {[k^2-M^2+i\epsilon][(k+p_i)^2-m^2+i\epsilon][(k+p_f)^2-m^2+i\epsilon]}, \\
P^{\mu}I_{{\pi}NN}^{(P)}(q^2) &=&
   i\int\frac{d^nk}{(2\pi)^n}\frac{k^{\mu}}
   {[k^2-M^2+i\epsilon][(k+p_i)^2-m^2+i\epsilon][(k+p_f)^2-m^2+i\epsilon]}, \\
\lefteqn{g^{\mu\nu}I_{{\pi}NN}^{(00)}(q^2)+P^{\mu}P^{\nu}I_{{\pi}NN}^{(PP)}(q^2)
   +q^{\mu}q^{\nu}I_{{\pi}NN}^{(qq)}(q^2)} \\
   &=&i\int\frac{d^nk}{(2\pi)^n}\frac{k^{\mu}k^{\nu}}
   {[k^2-M^2+i\epsilon][(k+p_i)^2-m^2+i\epsilon][(k+p_f)^2-m^2+i\epsilon]}.
\end{eqnarray*}

   The tensorial loop integrals can be reduced to scalar ones \cite{Passarino:1978jh} and
we obtain
\begin{eqnarray*}
I_{{\pi}N}^{(p)}(p^2) &=& \frac{1}{2p^2}
   \left[I_{\pi}-I_N-(p^2-m^2+M^2)I_{{\pi}N}(p^2)\right], \\
I_{{\pi}N}^{(00)}(p^2) &=& \frac{1}{2(n-1)}\left[I_N +2 M^2
   I_{{\pi}N}(p^2) +\frac{(p^2-m^2+M^2)}{p^2}I_{\pi N}^{(p)}(p^2)\right], \\
I_{{\pi}NN}^{(P)}(q^2) &=& \frac{1}{4m_N^2-q^2}
   \left[I_{{\pi}N}(m_N^2)-I_{NN}(q^2)-M^2I_{{\pi}NN}(q^2)\right], \\
I_{{\pi}NN}^{(00)}(q^2)&=&\frac{1}{n-2}\left\{\left[I_{{\pi}NN}(q^2)
   +I_{{\pi}NN}^{(P)}(q^2)\right]M^2+\frac{1}{2}I_{NN}(q^2)\right\},\\
I_{{\pi}NN}^{(PP)}(q^2)
   &=&\frac{1}{(n-2)(4m_N^2-q^2)}\left\{\left[(n-1)I_{{\pi}NN}^{(P)}(q^2)
   +I_{{\pi}NN}(q^2)\right]M^2\right. \\
   &&\left.-\frac{n-2}{2}I_{{\pi}N}^{(p)}(m_N^2)
   -\frac{n-3}{2}I_{NN}(q^2)\right\}, \\
I_{{\pi}NN}^{(qq)}(q^2) &=&
   -\frac{1}{(n-2)q^2}\left\{\left[I_{{\pi}NN}^{(P)}(q^2)+I_{{\pi}NN}(q^2)\right]M^2
   +\frac{n-2}{2}I_{{\pi}N}^{(p)}(m_N^2)+\frac{1}{2}I_{NN}(q^2)\right\}. \\
\end{eqnarray*}

Defining
\begin{displaymath}
\bar\lambda ={m^{n-4}\over 16\pi^2}\left\{ {1\over n-4}-{1\over 2}
\left[ \ln (4\pi) +\Gamma '(1)+1\right]\right\},
\end{displaymath}
and
\begin{displaymath}
\Omega=\frac{p^2-m^2-M^2}{2mM},
\end{displaymath}
the scalar loop integrals are given by \cite{Fuchs:2003qc}
\begin{displaymath}
I_\pi=2M^2\bar{\lambda}+\frac{M^2}{8\pi^2}\ln\left(\frac{M}{m}\right),
\end{displaymath}
\begin{displaymath}
I_N=2m^2\bar{\lambda},
\end{displaymath}
\begin{displaymath}
I_{\pi\pi}(q^2)=2\bar{\lambda}+\frac{1}{16\pi^2}\left[ 1 +
2\ln\left(\frac{M}{m}\right)
                    + J^{(0)}\left(\frac{q^2}{M^2}\right)\right],
\end{displaymath}
\begin{displaymath}
I_{NN}(q^2)=2\bar{\lambda}+\frac{1}{16\pi^2}\left[1+J^{(0)}
\left(\frac{q^2}{m^2}\right)\right]
\end{displaymath}
\begin{displaymath}
I_{\pi N}(p^2)=2\bar{\lambda}+\frac{1}{16\pi^2}\left[-1
+\frac{p^2-m^2+M^2}{p^2}\ln\left(\frac{M}{m}\right)
+\frac{2mM}{p^2}F(\Omega)\right],
\end{displaymath}
where
\begin{eqnarray*}
J^{(0)}(x)
&=&\int_0^1 dz \ln[1+x(z^2-z)-i\epsilon]\\
&=&
 \left \{ \begin{array}{ll}
-2-\sigma\ln\left(\frac{\sigma-1}{\sigma+1}\right),&x<0,\\
-2+2\sqrt{\frac{4}{x}-1}\,\mbox{arccot}
\left(\sqrt{\frac{4}{x}-1}\right),&0\le x<4,\\
-2-\sigma\ln\left(\frac{1-\sigma}{1+\sigma}\right)-i\pi\sigma,&
4<x,
\end{array} \right.
\end{eqnarray*}
with
\begin{displaymath}
\sigma(x)=\sqrt{1-\frac{4}{x}},\quad x\notin [0,4],
\end{displaymath}
and
\begin{eqnarray*}
F(\Omega) &=& \left \{ \begin{array}{ll}
\sqrt{\Omega^2-1}\ln\left(-\Omega-\sqrt{\Omega^2-1}\right),&\Omega\leq -1,\\
\sqrt{1-\Omega^2}\arccos(-\Omega),&-1\leq\Omega\leq 1,\\
\sqrt{\Omega^2-1}\ln\left(\Omega+\sqrt{\Omega^2-1}\right)
-i\pi\sqrt{\Omega^2-1},&1\leq \Omega.
\end{array} \right.
\end{eqnarray*}
   Integrals with three propagators were analyzed numerically
using a Schwinger parametrization.

   For purely mesonic integrals only the terms proportional to
$\bar{\lambda}$ have to be subtracted.
   To determine the infrared regular parts $R$ of the scalar loop
integrals, we use the method described in \cite{Schindler:2003xv}.
   On-shell-kinematics are assumed for the subtraction terms.
   Note that we also list divergent terms, as they might give
finite contributions in the expressions for tensor integrals.
\begin{eqnarray*}
  R_N &=& I_N, \\
  R_{NN} &=& I_{NN},\\
  R_{\pi N} &=& \bar\lambda\left[2-\frac{M^2}{m^2}\left(1-8c_1m\right)
+\frac{3 g_A^2 M^3}{16 \pi F^2 m}\right] -\frac{1}{16\pi^2}-\frac{M^2}{32\pi^2m^2}\left(3+8c_1m\right)\\
  && -\frac{3 g_A^2 M^3}{512 \pi^3 F^2 m}+{\cal O}(p^4), \\
  R_{\pi NN} &=& \frac{\bar\lambda}{m^2}\left[1 +\frac{q^2}{6m^2}+8c_1\frac{M^2}{m}
+\frac{3 g_A^2 M^3}{16 \pi F^2 m}\right] +\frac{1}{32\pi^2m^2}-\frac{M^2}{32\pi^2m^4}\left(1-16c_1m\right)\\
  && +\frac{3 g_A^2 M^3}{256 \pi^3 F^2 m^3}+{\cal O}(p^4).
\end{eqnarray*}

\end{document}